\documentclass[12pt,preprint]{aastex}
\shorttitle{Cepheid Pulsations}
\shortauthors{Lane et al.}

\begin{document}

\title{Long Baseline Interferometric Observations of Cepheids.}

\author{Benjamin\,F. Lane} 
\affil{Department of Geological \& Planetary Sciences, MS
       150-21, California Institute of Technology, Pasadena CA 91125,
       U.S.A.}
\email{ben@gps.caltech.edu}

\author{Michelle J. Creech-Eakman}
\affil{Caltech/JPL Postdoctoral Scholar, Jet Propulsion Laboratory, California Institute of Technology, 4800 Oak Grove Drive, Pasadena, CA 91109, U.S.A. }
\email{mce@huey.jpl.nasa.gov}
\and
\author{Tyler E. Nordgren}
\affil{University of Redlands, Department of Physics, 1200 E Colton Ave, Redlands, California, U.S.A.}
\email{Tyler\_Nordgren@redlands.edu}

\begin{abstract}

We present observations of the galactic cepheids $\eta$ Aql and
$\zeta$ Gem.  Our observations are able to resolve the diameter
changes associated with pulsation. This allows us to determine the
distance to the Cepheids independent of photometric observations. We
determine a distance to $\eta$ Aql of $ 320\pm32$ pc, and a distance 
to $\zeta$ Gem of $ 362\pm38$ pc. These observations
allow us to calibrate surface brightness relations for use in
extra-galactic distance determination. They also provide
a measurement of the mean diameter of these Cepheids, which is useful
in constructing structural models of this class of star.
\end{abstract}


\section{Introduction}

The class of pulsating stars known as Cepheids is a cornerstone in
determining the distances to nearby galaxies. This is because Cepheids
exhibit a well-behaved period-luminosity relation which can be locally
calibrated \citep{jacoby92}. In addition, these stars are massive and
thus intrinsically very luminous, making it possible to observe
Cepheids located in very distant galaxies \citep{tan99,feast99}.
Because of the usefulness and fundamental importance of Cepheids, it
is important to calibrate their period-luminosity relation. This has
been done using a variety of methods, including parallax \citep{esa97,
fc97}, Baade-Wesselink methods \citep{wesselink46,bersier97} and
surface brightness \citep{laney95,fg97,ripepi97}. The
period-luminosity relations used currently have uncertainties on the
order of 0.09 mag \citep{feast99}, which in turn make up a significant
portion of the systematic uncertainty in estimates to the Large
Magellanic Cloud.

Using long-baseline stellar interferometry it is possible to resolve
the diameter changes undergone by a nearby Cepheid during a
pulsational cycle. When such diameter measurements are combined with
radial velocity measurements of the stellar photosphere, it is
possible to determine the size of and distance to the Cepheid. Such a
direct measurement is independent of photometric observations and
their associated uncertainties.

The Palomar Testbed Interferometer (PTI) is located on Palomar
Mountain near San Diego, CA \citep{colavita99}. It combines starlight
from two 40-cm apertures to measure the amplitude (a.k.a. visibility)
of the resulting interference fringes. There are two available
baselines, one 110-m baseline oriented roughly North-South (hereafter
N-S), and one 85-m baseline oriented roughly North-Southwest (called
N-W).  In a previous paper \citep{lane00} we presented observations using PTI
of the Cepheid $\zeta$ Gem.  Here we report on additional
interferometric observations of $\zeta$ Gem, as well as a second
Galactic Cepheid, $\eta$ Aql. These observations allow us to determine
the distances to these Cepheids with the aim of reducing the
uncertainty in currently used period-luminosity relations for
Cepheids.

\section{Observations}

We observed the nearby galactic cepheids $\eta$ Aql and $\zeta$ Gem on
22 nights between 2001 March 13 and 2001 July 26. The observing
procedure followed standard PTI practice
\citep{boden98,colavita99}. For the observations of $\eta$ Aql the N-W
baseline was used, while observations of $\zeta$ Gem used the N-S
baseline. Each nightly observation consisted of approximately ten
130-second integrations during which the fringe visibility was
averaged.  The measurements were done in the $1.52-1.74~\mu$m
(effective central wavelength $1.65~\mu$m) wavelength region, similar
to the astronomical $H$ band.  Observations of calibration sources
were rapidly (within less than $\sim 10$ minutes) interleaved with the
Cepheid observations, and after each 130-second integration the
apertures were pointed to dark sky and a 30-second measurement of the
background light level was made.

The calibrators were selected to be located no more than 16 degrees
from the primary target on the sky and to have similar $H$-band
magnitudes. In choosing calibration sources we avoided known binary or
highly variable stars. The calibrators used are listed in Table
\ref{tab:calibs}. In this paper we make use of previously published
observations of the Cepheid $\zeta$ Gem \citep{lane00}. However, in
order to improve on the previously published results we carried out
additional observations of this source on 2001 March 13--15. We also
observed additional unresolved calibrators in order to reduce the
level of systematic uncertainty. The original data have been jointly
re-reduced using the improved calibrator diameters and
uncertainties. However, note that the primary calibrator diameter has
not changed from the value used in \citet{lane00}.

\section{Analysis \& Results}

\subsection{Fringe Visibilities \& Limb Darkening}
PTI uses either a 10 or 20 ms sample rate.  Each such sample provides
a measure of the instantaneous fringe visibility and phase. While the
phase value is converted to distance and fed back to the active delay
line to provide active fringe tracking, the measured fringe visibility
is averaged over the entire 130-second integration. The statistical
uncertainty in each measurement is estimated by breaking the 130 second
integration into five equal-time segments and measuring the standard
deviation about the mean value.

The theoretical relation between source brightness distribution and
fringe visibility is given by the van Cittert-Zernike theorem. For a
uniform intensity disk model the normalized fringe visibility (squared) can be
related to the apparent angular diameter as
\begin{equation}
V^2 = \left( \frac{2 \; J_{1}(\pi B \theta_{UD} / \lambda_0)}{\pi B\theta_{UD} / \lambda_0} \right)^2
\label{eq:V2_single}
\end{equation}
where $J_{1}$ is the first-order Bessel function, $B$ is the projected
aperture separation, $\theta_{UD}$ is the apparent angular diameter of
the star in the uniform-disk model, and $\lambda_0$ is the center-band
wavelength of the observation. It follows that the fringe visibility of a 
point source measured by an ideal interferometer should be unity. 
For a more realistic model that includes limb darkening one can derive a 
conversion factor between a uniform-disk diameter ($\theta_{UD}$) and 
a limb-darkened disk diameter ($\theta_{LD}$) given by \citep{welch94}
\begin{equation}
\theta_{UD} = \theta_{LD} \sqrt{1 - \frac{A}{3} - \frac{B}{6}}
\end{equation}
where A and B are quadratic limb darkening coefficients, determined by
the spectral type of the source \citep{claret95}.  The limb darkening
correction factors ($k = \theta_{UD}/\theta_{LD}$) used for the 
Cepheids are shown in Table \ref{tab:targ} and for the 
calibrators in Table \ref{tab:calibs}.

\subsection{Visibility Calibration}

The first step in calibrating visibilities measured by PTI is to
correct for the effects of detector background and read-noise, the
details of which are discussed in \citet{colavita99} and
\citet{colavita99b}.  However, the visibilities thus produced are not
yet final: due to a variety of effects, including systematic
instrumental effects, intensity mismatches, and atmospheric
turbulence, the fringe visibility of a source measured by PTI is lower
than that predicted by Eq. 1. In practice the system response function
(called the system visibility) is typically $\sim 0.75$ and
furthermore is variable on 30 minute timescales. Hence the
visibilities must be calibrated by observing sources of known
diameter.

Determining the diameter of the calibration sources was a multi-step
process in which we made use of both models and prior observations.
For each Cepheid we designated a single, bright K giant as a primary
calibrator, which was always observed in close conjunction with the
target Cepheid (HD 189695 for $\eta$ Aql, and HD 49968 for $\zeta$
Gem). We used model diameter estimates for the primary calibrators 
from previously published results based on spectro-photometry and
modeling \citep{cohen99}.

In order to verify that the primary calibrators were stable and had
angular diameters consistent with the \citet{cohen99} results, we
observed them together with a number of secondary calibrators. These
secondary calibrators were typically less resolved than the primary
calibrators and hence less sensitive to uncertainties in their
expected angular diameter. However, they were fainter than the primary
calibrators, and tended to be located further away on the sky. For the
secondary calibrators an apparent diameter was estimated using three
methods: (1) we used available archival photometry to fit a black-body
model by adjusting the apparent angular diameter, bolometric flux and
effective temperature of the star in question so as to fit the
photometry. (2) We repeated the above fit while constraining the
effective temperature to the value expected based on the published
spectral type.  (3) We estimated the angular diameter of the star
based on expected physical size (derived from spectral type) and
distance (determined by Hipparcos). We adopted the weighted (by the
uncertainty in each determination) mean of the results from the above
methods as the final model diameter for the secondary calibrators, and
the uncertainty in the model diameter was taken to be the deviation
about the mean.

In addition to the model-based diameter estimates derived above we
also used extensive interferometric visibility measurements for the
primary and secondary calibrators; given that several of the
calibrators were observed within a short enough period of time that
the system visibility could be treated as constant, it was possible to
find a set of assumed calibrator diameters that are maximally
self-consistent, by comparing observed diameter ratios for which the
system visibility drops out. To illustrate, let $\theta_i$ be an
adjustable parameter, representing the diameter of star $i$. Let
$\hat{\theta}_i$ and $\sigma_{\hat{\theta_i}}$ be the theoretical
model diameter and uncertainty for star $i$ derived above, and let
$\tilde{R_{ij}}$ and $\sigma_{\tilde{R_{ij}}}$ be the
interferometrically observed diameter ratio and uncertainty of stars
$i$ and $j$. For notational simplicity, define $R_{ij}$ as the ratio
of $\theta_i$ and $\theta_j$.  Define the quantity
\begin{equation}
 \chi^2 = \sum_{i} \left[ \frac{ \hat{\theta}_i - \theta_i }{\sigma_{\hat{\theta_i}}} \right]^2 \
+ \sum_{i} \sum_{j < i} \left[ \frac{ \tilde{R_{ij}} - R_{ij}}{\sigma_{\tilde{R_{ij}}}} \right]^2
\end{equation}
By adjusting the set of $\theta_i$ to minimize $\chi^2$ we produce a
set of consistent calibrator diameters, taking into account both input
model knowledge and observations. The resulting diameter values are
listed in Table \ref{tab:calibs}. Uncertainties were estimated using
the procedure outlined in \citet{press86} assuming normally
distributed errors.

We verified that the primary calibrators were stable as follows: using
the secondary calibrators to calibrate all observations of the primary
calibrators we fit a constant-diameter, single-star, uniform-disk
model to the primary calibrators. In all cases the scatter about the
single-star model was similar to expected system performance
\citep{boden98}: for HD 189695, 21 points were fit, the average
deviation in $V^2$ was 0.035 and the goodness-of-fit parameter of the
line fit, $\chi^2$ per degree of freedom ($\chi^2_{dof}$, not to be
confused with Eq. 3 above), in the line fit was 0.46. For HD 49968, 82
points were fit, the average deviation was 0.038 and $\chi^2_{dof}$ =
0.76.

\begin{table}
\begin{center}
\begin{tabular}{ccccccc}
	   &              &            &  &              &                & \\
\tableline 
\tableline                
Star       & Alternate   & Period   & Epoch        &   Limb Dark.  \\
Name	   & Name	 & (d)      & JD          &   Factor (k)  \\
\tableline
$\eta$ Aql & HD 187929   & 7.176711 & 2443368.962  & $0.97 \pm 0.01$  \\
$\zeta$ Gem& HD 52973    & 10.150079& 2444932.736  & $0.96 \pm 0.01$\\
\tableline
\end{tabular}
\caption{\label{tab:targ} Relevant parameters of the Cepheids. The limb darkening 
factor is defined as $k = \theta_{UD}/\theta_{LD}$.}

\end{center}
\end{table}

\begin{table}
\begin{center}
\begin{tabular}{lcccccccc}
	   &              &              &              &                & & & & \\
\tableline 
\tableline                
Calibrator & Spectral    & Diameter Used      &   Limb Dark.    & Used to   & Cal. & Angular Sep. \\
	   & Type	 & $\theta_{UD}$ (mas) &   Factor (k)    & calibrate & Type   & (deg) \\
\tableline
HD 189695  & K5 III      & $1.89 \pm 0.07$   & $0.943\pm0.007$ & $\eta$ Aql & Pri. Cal & 7.8  \\
HD 188310  & G9.5 IIIb   & $1.57 \pm 0.08$   & $0.955\pm0.007$ & $\eta$ Aql & Sec. Cal & 8.2  \\ 
HD 181440  & B9 III      & $0.44 \pm 0.05$   & $0.975\pm0.007$ & $\eta$ Aql & Sec. Cal & 7.5 \\	
HD 49968   & K5 III      & $1.78 \pm 0.02$   & $0.939\pm0.006$ & $\zeta$ Gem & Pri. Cal & 4.1  \\
HD 48450   & K4 III      & $1.94 \pm 0.02 $  & $0.949\pm0.007$ & $\zeta$ Gem & Sec. Cal & 9.5 \\
HD 39587   & G0 V        & $1.09 \pm 0.04$   & $0.963\pm0.006$ & $\zeta$ Gem & Sec. Cal & 16  \\
HD 52711   & G4 V        & $0.55 \pm 0.04$   & $0.962\pm0.006$ & $\zeta$ Gem & Sec. Cal & 8.8 \\

\tableline
\end{tabular}
\caption{\label{tab:calibs} Relevant parameters of the calibrators. The angular separation 
listed is the angular distance from the calibrator to the Cepheid it is used to calibrate. }

\end{center}
\end{table}

While analyzing the data it was noticed that during observations with
the N--W baseline of relatively low declination sources, such as
$\eta$ Aql and its calibrators, the stability of the interferometer
system visibility was strongly dependent on the hour angle of the
source: for observations of $\eta$ Aql obtained at positive hour
angles the scatter in the system visibility increased by a factor of
2--3, while the mean value trended down by 20\%/hr. There are two
potential explanations for this effect: (1) for these observations the
optical delay lines are close to their maximum range, which can
exacerbate internal system misalignments and lead to vignetting. (2)
When observing low declination sources past transit, the siderostat
orientation is such that surface damage near the edge of one of the
siderostat mirrors causes vignetting. Thus it was decided to discard
observations of $\eta$ Aql taken at positive hour angles,
corresponding to $\sim 20$\% of the available data. We note that
including the data does not significantly change the final results
($\sim 0.3\sigma$), it
merely increases the scatter substantially (for the pulsation fit
discussed below the goodness-of-fit parameter $\chi^2_{dof}$ increased
from 1.06 to 4.5).
 
\subsection{ Apparent Angular Diameter}

Once the measured visibilities were calibrated we used all the
available calibrated data from a given night to determine the apparent
uniform-disk angular diameter of the target Cepheid on that particular night
by fitting to a model given by Eq. 1.  Results are given in
Tables \ref{tab:diams_ea} and \ref{tab:diams_zg} and plotted in
Fig. 1. Uncertainties were estimated based on the scatter about the
best fit. It should be noted that although $\eta$ Aql is known to have
a companion \citep{bv85} it is sufficiently faint (average $\Delta m_H
= 5.75$ mag) that it will have a negligible effect ($\Delta V^2 \sim
0.005$) on the fringe visibilities measured in the H band.

\begin{table}
\begin{center}
\begin{tabular}{ccc}
	   &                                               \\
\tableline 
\tableline                
Epoch        &  Angular Diameter & No. Scans  \\
JD-2400000.5 &  $\theta_{UD}$ (mas) &  \\
\tableline
52065.420 & $    1.654 \pm 0.011$ & 9 \\ 
52066.414 & $    1.654 \pm 0.017$ & 9 \\ 
52067.405 & $    1.694 \pm 0.040$ & 8 \\ 
52075.383 & $    1.740 \pm 0.027$ & 12 \\ 
52076.384 & $    1.799 \pm 0.014$ & 9 \\ 
52077.372 & $    1.822 \pm 0.021$ & 13 \\ 
52089.350 & $    1.715 \pm 0.019$ & 11 \\ 
52090.354 & $    1.798 \pm 0.020$ & 9 \\ 
52091.346 & $    1.764 \pm 0.022$ & 7 \\ 
52095.360 & $    1.567 \pm 0.049$ & 1 \\ 
52099.337 & $    1.800 \pm 0.025$ & 2 \\ 
52101.329 & $    1.632 \pm 0.037$ & 5 \\ 
52103.293 & $    1.656 \pm 0.040$ & 7 \\ 
52105.300 & $    1.798 \pm 0.024$ & 6 \\ 
52106.283 & $    1.816 \pm 0.016$ & 19 \\ 
52107.302 & $    1.809 \pm 0.027$ & 11 \\ 
52108.308 & $    1.702 \pm 0.032$ & 7 \\ 
52116.276 & $    1.611 \pm 0.023$ & 7 \\ 

\tableline
\end{tabular}
\caption{\label{tab:diams_ea} The measured uniform-disk diameters of $\eta$ Aql.
The uncertainties are the statistical uncertainty 
from the scatter during a night, and do not include systematic uncertainty 
in the calibrator diameters; this adds an additional uncertainty of $0.07$ mas in the aggregate mean diameter. }

\end{center}
\end{table}

\begin{table}
\begin{center}
\begin{tabular}{ccc}
	   &                                               \\
\tableline 
\tableline                
Epoch        &  Angular Diameter    & No. Scans\\
JD-2400000.5 &  $\theta_{UD}$ (mas) & \\
\tableline
51605.226 & $    1.676 \pm 0.015$ & 15 \\ 
51606.241 & $    1.675 \pm 0.047$ & 3 \\ 
51614.192 & $    1.797 \pm 0.060$ & 7 \\ 
51615.180 & $    1.737 \pm 0.031$ & 10 \\ 
51617.167 & $    1.587 \pm 0.028$ & 10 \\ 
51618.143 & $    1.534 \pm 0.008$ & 11 \\ 
51619.168 & $    1.549 \pm 0.018$ & 15 \\ 
51620.169 & $    1.585 \pm 0.028$ & 15 \\ 
51622.198 & $    1.673 \pm 0.046$ & 6 \\ 
51643.161 & $    1.663 \pm 0.012$ & 9 \\ 
51981.182 & $    1.685 \pm 0.014$ & 23 \\ 
51982.164 & $    1.636 \pm 0.020$ & 16 \\ 
51983.201 & $    1.589 \pm 0.021$ & 15 \\ 
51894.387 & $    1.619 \pm 0.019$ & 13 \\ 
51895.369 & $    1.629 \pm 0.014$ & 12 \\ 

\tableline
\end{tabular}
\caption{\label{tab:diams_zg} The measured uniform-disk diameters of $\zeta$ Gem.
The uncertainties are the statistical uncertainty 
from the scatter during a night, and do not include systematic uncertainty 
in the calibrator diameters; this adds an additional uncertainty of $0.024$ mas 
in the aggregate mean diameter.
}

\end{center}
\end{table}

\begin{figure}[ht]
\epsscale{0.8}
\plotone{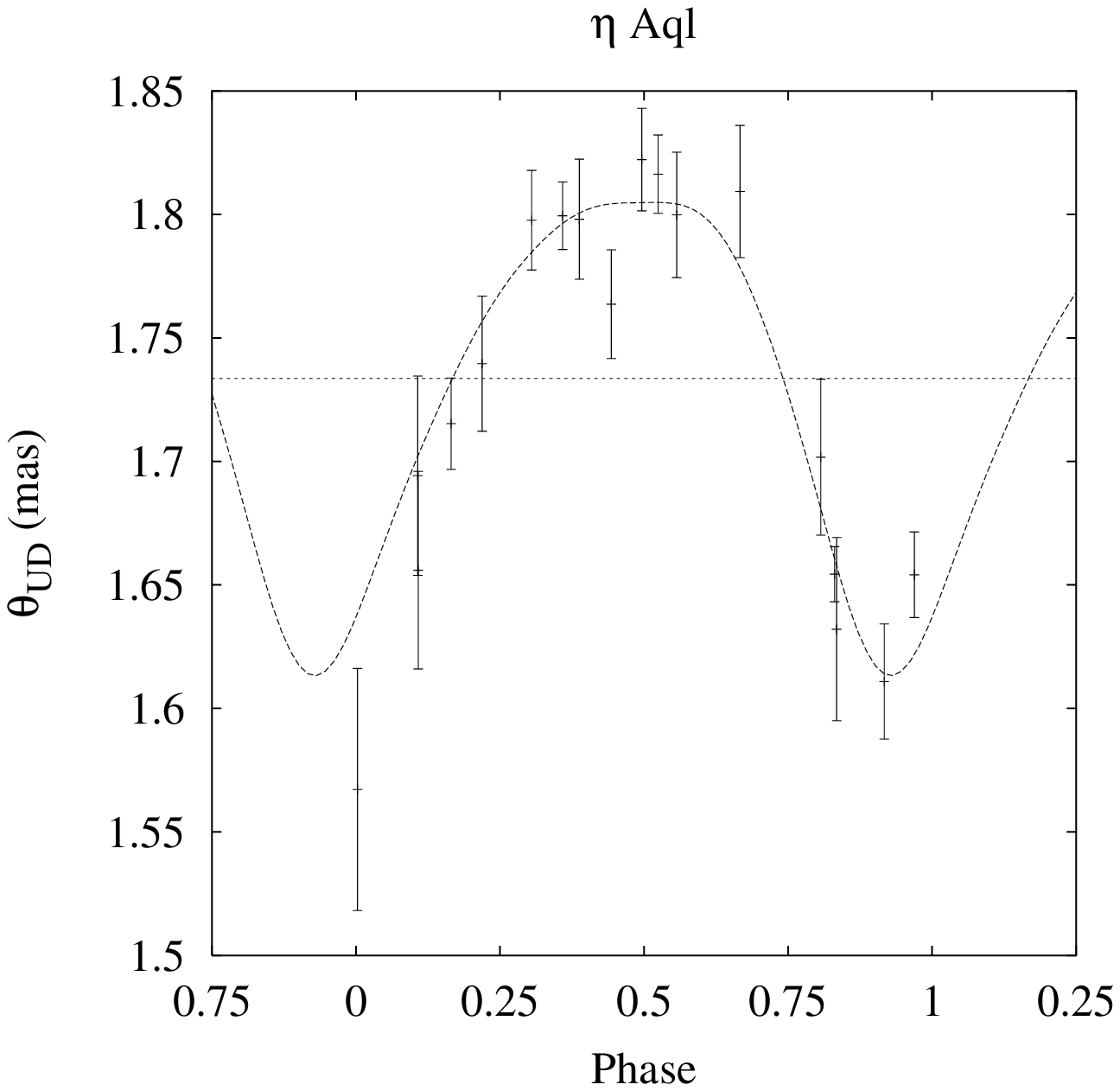}
\plotone{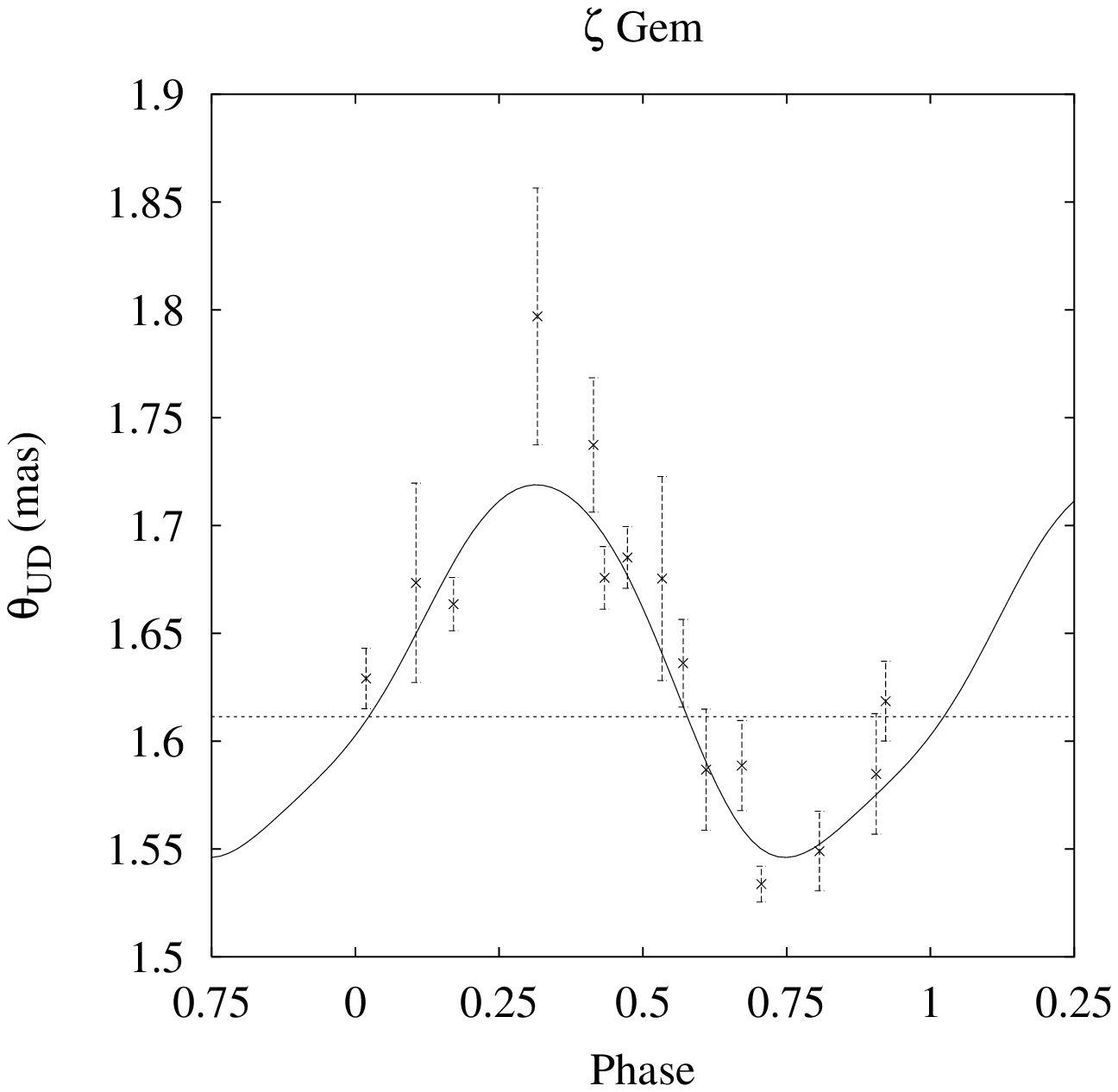}
\caption[Cepheid Angular Diameters]{\label{diams} The angular diameters of $\eta$ Aql (top) and
$\zeta$ Gem (bottom) as a function of pulsational phase, together with
a model based on radial velocity data, but fitting for distance, mean
radius and phase shift. Also shown is the result of fitting a line to
all the data. The fits are extended past phase 0 for clarity.}
\end{figure}

 It is clear from Fig. \ref{diams} that the measured angular diameters 
are not constant with time. Fitting a constant-diameter model to the data
produces a rather poor fit (see Table \ref{tab:result}). However, we 
list the resulting mean angular diameters in order to facilitate comparison 
with previous interferometric results.

\subsection{Distances \& Radii}

Determining the distance and radius of a Cepheid via the
Baade-Wesselink method requires comparing the measured changes in
angular diameter to the expansion of the Cepheid photosphere measured
using radial velocity techniques.  In order to determine the expansion
of the Cepheid photospheres we fit a fifth-order Fourier series to
previously published radial velocities. For $\eta$ Aql we used data
from \citet{bersier02} as well as data published by \citet{jw81,jw87},
while for $\zeta$ Gem we used data from \citet{bersier94}. Both sets
of data were from measurements made at optical wavelengths. The
measured radial velocities were converted to physical expansion rates
using a projection factor (p-factor), which depends on the detailed
atmospheric structure and limb darkening of the Cepheid as well as on the
details of the equipment and software used in the measurement
\citep{hind86,albrow94}. It is important to note that the p-factor is
not expected to stay constant during a pulsational cycle. The exact
phase dependence of the p-factor is beyond the scope of this
paper. However, for $\eta$ Aql and $\zeta$ Gem, the net effect of a
variable p-factor can be approximated by using a 6\% larger constant
p-factor \citep{sabbey95}.  Thus for both Cepheids we use an effective
p-factor of $1.43 \pm 0.06$, constant for all pulsational phases.

We convert the radial velocity Fourier series into a physical size
change by integrating and multiplying by limb-darkening and p-factors.
Although the limb-darkening does vary with changing ${\rm T}_{eff}$
during a pulsational cycle, the effect is small: for $\zeta$ Gem $k$
varies from 0.960 to 0.967, i.e.\ less than the quoted uncertainty.
The size change can in turn be converted into an angular size model
with three free parameters: the mean physical radius, the distance to
the star, and a phase shift. The latter is to account for
possible period changes, inaccuracies in period or epoch, or phase
lags due to level effects (where the optical and infrared photospheres
are at different atmospheric depths; see below).  We adjust the model
phase, radius and distance to fit the observed angular diameters.
Results of the fits for $\eta$ Aql and $\zeta$ Gem are given in Table
\ref{tab:result}.

There are several sources of uncertainty in the above fits: in
addition to the purely statistical uncertainty there are systematic
uncertainties of comparable magnitude. The three primary sources of
systematic uncertainty are: uncertainty in the calibrator diameters,
uncertainty in the p-factor, and uncertainty in the limb darkening
coefficients. The magnitude of each effect was estimated separately by
re-fitting the model while varying by $\pm 1\sigma$ each relevant
parameter separately. The total systematic uncertainty was calculated
as 
\begin{equation}
\sigma_{sys}^2 = \sigma_{cal}^2 + \sigma_{p-fac}^2 + \sigma_{limb dark.}^2 
\end{equation}

In order to explore the possibility of wavelength-dependent effects on
the measured radial velocity, e.g.\ due to velocity gradients in the
Cepheid atmospheres (``level effects''), we re-fit for the radius and
distance of $\eta$ Aql using a radial velocity curve based on radial
velocity data obtained at wavelengths of $1.1$ and $1.6 \mu$m by
\citet{sasselov90}. Because of the limited number of observations
available (e.g only 3 $H$-band measurements of $\eta$ Aql) we used the
shape of the radial velocity curve derived from the fit to the optical
data (i.e.\ by using the same Fourier coefficients); the IR data was
only used to determine an overall amplitude of the velocity curve. For
the IR points we used an effective p-factor of $1.41 \pm 0.03$ as
recommended by D. Sasselov (private communication) and based on an
analysis by \citet{sabbey95}, taking into account both the use of a
constant p-factor and the use of parabolic line fitting. The resulting
best-fit parameters are very similar to those based on optical radial
velocities (i.e. Table \ref{tab:result}): ${\rm D} = 333 \pm 30$ pc
and ${\rm R}=64.2 \pm 6 {\rm R}_{\odot}$. A similar fit for $\zeta$
Gem gives ${\rm D} = 359 \pm 37$ pc and ${\rm R}=62.2 \pm 5.7 {\rm
R}_{\odot}$. Hence we conclude that the effects of wavelength
dependence of the radial velocity are at present smaller than other
sources of uncertainty.

\begin{table}
\begin{center}
\begin{tabular}{cccc}
	   &                       &               &                         \\
\tableline 
\tableline                
   Cepheid &  Fit Type             & Parameter     & Best-Fit Results        \\
	   &                       &     	   & Value $\pm$ $\sigma_{Tot}$ ($\sigma_{Stat.}/\sigma_{Sys.}$)   \\
\tableline
$\eta$ Aql & Pulsation Fit         & Distance (D)  &  $ 320\pm32$ (24/21) pc \\
           & No. Pts. = 18         & Radius (R)    &  $ 61.8\pm7.6$ (4.5/6.1) ${\rm R}_{\odot}$ \\
           & $\chi^2_{dof}$ = 1.06   & Phase ($\phi$)&  $ 0.02\pm0.011$ ($0.011/5 \times 10^{-4}$) cycles\\ 
\cline{2-4} 
	   & Line Fit              & $\theta_{UD}$ & $ 1.734\pm 0.070 (0.018/0.068)$ mas  \\
	   & $\chi^2_{dof}$ = 13.4 &               &                        \\

\tableline
$\zeta$ Gem & Pulsation Fit        & Distance (D)          &  $ 362\pm38$ (35/15) pc                \\
            &No. Pts. = 15         & Radius (R)            &  $ 66.7\pm7.2$ (6.3/3.4) ${\rm R}_{\odot}$ \\
            &$\chi^2_{dof}$ = 1.82   & Phase ($\phi$)        &  $ 0.013\pm0.016$ ($0.016/3 \times 10^{-5}$) cycles            \\
\cline{2-4} 
	   & Line Fit              & $\theta_{UD}$ & $ 1.613\pm 0.029 (0.017/0.024)$ mas  \\
	   & $\chi^2_{dof}$ = 14.6 &               &                       \\

\tableline

\end{tabular}
\caption{\label{tab:result} Best-fit Cepheid parameters and their
uncertainties, as well as mean apparent uniform-disk angular diameter
($\theta_{UD}$) determined from fitting a line to all of the data. The
uncertainties of the best-fit parameters are broken down into
statistical ($\sigma_{Stat.}$) and systematic ($\sigma_{Sys.}$)
uncertainties. The goodness-of-fit parameter is a weighted $\chi^2$
divided by the number of degrees of freedom ($\chi^2_{dof}$) in the fit.  The
$\chi^2_{dof}$ of the fits are calculated from data that does not have
the systematic (calibrator) uncertainty folded in since it applies 
equally to all points. }

\end{center}
\end{table}

\begin{table}
\begin{center}
\begin{tabular}{clccc}
	   &                       &               &                     &    \\
\tableline 
\tableline  
Cepheid	   & Reference             & Radius             & Distance   & Angular Diameter     \\
           &                       &  ${\rm R}_{\odot}$ & (pc)       &  $\theta_{LD}$ (mas) \\
\tableline 
\tableline 
$\eta$ Aql  &  this work       &  $61.8\pm7.6$     &  $ 320\pm32$        &  $ 1.793\pm 0.070$  \\
	    &\cite{nordgren00} &                   &                     &  $ 1.69\pm 0.04$   \\           
            &\citet{ripepi97}  &  $57\pm3$         &                     &    \\ 
            &  \citet{esa97}   &                   &  $360^{+174}_{-89}$ &    \\
            &\citet{sasselov90}&  $62\pm6$         &                     &    \\
	    &\cite{fern89}     &  $53\pm5$         &  $275 \pm 28$       &    \\
            &\citet{mb87}      &  $55\pm4$         &                     &    \\   	   
\tableline   
$\zeta$ Gem &  this work        &    $ 66.7\pm7.2$  &  $ 362\pm38$       &  $ 1.675\pm 0.029$  \\
	    & \citet{lane00}    & $ 62\pm11$        &  $ 336\pm44$       &  $ 1.62\pm 0.3$  \\
            & \citet{ker01}     &                   &                    &  $ 1.69^{+0.14}_{-0.16}$  \\
	    & \cite{nordgren00} &                   &                    &  $ 1.55\pm 0.09$    \\
 	    & \citet{esa97}     &                   & $358^{+147}_{-81}$ &    \\
            & \citet{ripepi97}  & $86\pm4$          &                    &    \\
            & \citet{bersier97} & $89.5\pm13$       & $498\pm84$         &    \\
	    & \citet{ksn97}     & $69.1^{+5.5}_{-4.8}$&                   &    \\
	    & \citet{sabbey95}  & $64.4 \pm 3.6$    &                    &     \\
            & \citet{mb87}      & $65\pm12$         &                    &    \\
\tableline    
\end{tabular}
\caption{\label{tab:comparison} A comparison between the various available radius, distance and angular size 
determinations. The \citet{nordgren00} results are based on $R$ band ($740$ nm) observations, while the 
\citet{ker01} result is in the $K$ band ($2.2~\mu$m).
}
\end{center}
\end{table}

 The derived parameters (mean radius, distance and mean uniform-disk
angular diameter) can be compared to previously published values,
derived using a range of techniques (see Table \ref{tab:comparison}),
including parallax and a variety of surface brightness
techniques. There are also several interferometric diameter
measurements available in the literature, although to date no other
interferometers have directly resolved Cepheid pulsations. Thus,
directly measured angular diameters can only be compared in a
phase-averaged sense. 

\subsection{Surface Brightness Relations}

A wide variety of Cepheid surface brightness relations have been used 
by various authors \citep{be76,laney95,fg97} to derive Cepheid distance scales. 
We define as surface brightness the quantity
\begin{equation}
\label{eqn:sb}
F_{i} = 4.2207 - 0.1 m_i - 0.5 \log(\theta_{LD})
\end{equation}
where $F_{i}$ is the surface brightness in magnitudes in
passband $i$, $m_i$ is the apparent magnitude in that band, and
$\theta_{LD}$ is the apparent angular diameter of the star. With the
above relation and a good estimate of $F_{i}$ one can determine
the angular diameter based on photometry alone. Conversely, given
measured angular diameters and multi-band photometry it is possible to
calibrate $F_{i}$ by finding a simple (e.g.\ linear) relation
between $F_{i}$ and a variety of color indices (e.g $V-K$). We
define the following relations
\begin{equation}
  F_{V,1} = a + b (V-K)
\end{equation}
\begin{equation}
  F_{V,2} = a + c (V-R)
\end{equation}
Note that
consistency requires a common zero-point (cf. an A0V star where
$(V-R)=(V-K)=0$).

\begin{figure}[ht]
\epsscale{0.8}
\plotone{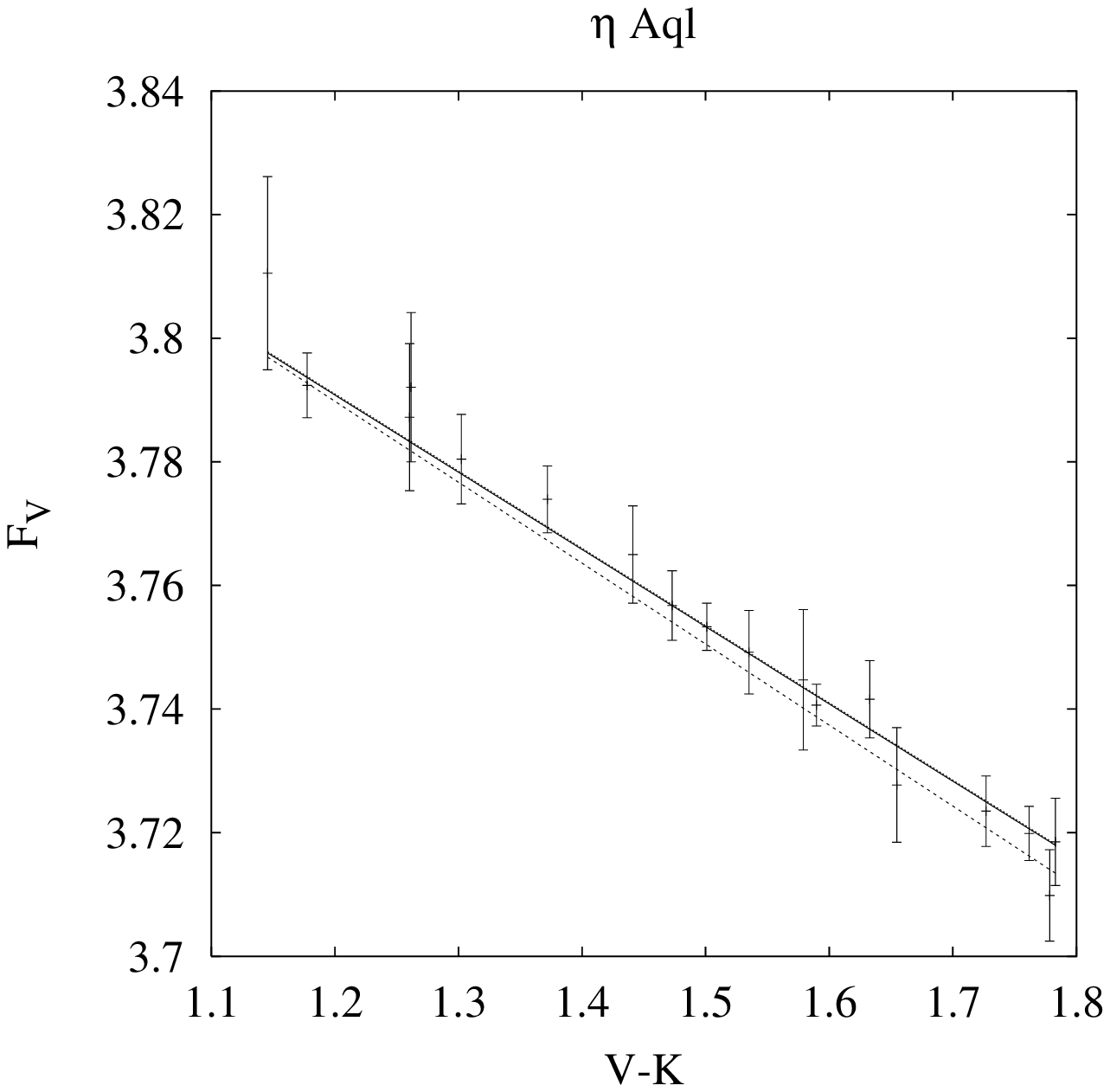}
\plotone{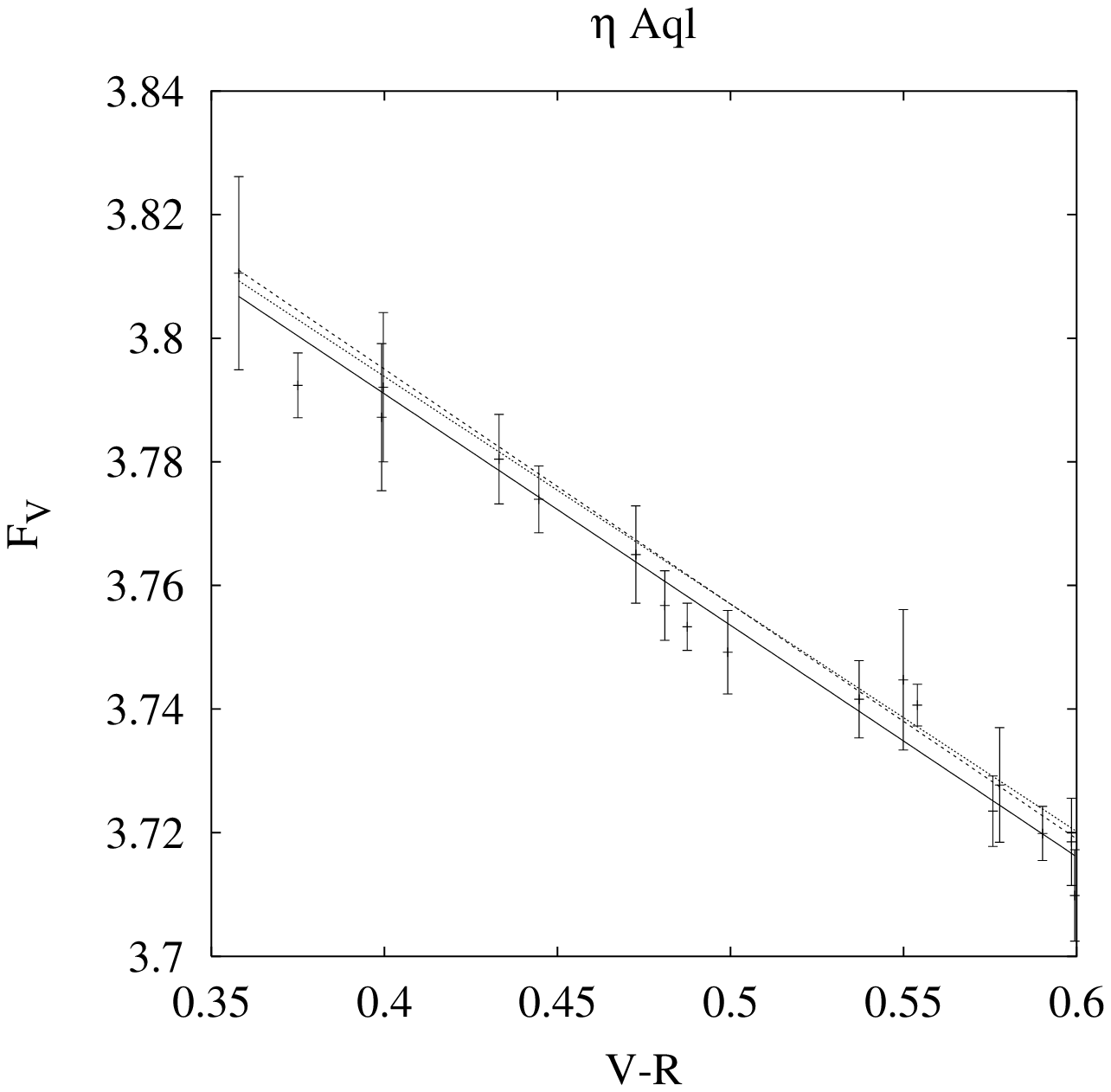}
\caption[Cepheid Surface Brightness]{\label{fig:fv} Dereddened ${\rm F}_V$ vs. $V-K$ (top) and
$V-R$ (bottom) for $\eta$ Aql. The solid line is the weighted linear
least-squares fit to the data. The dashed line represents the relation
from \citet{fg97}, and the dotted line represents the \citet{nordgren01}
result. }
\end{figure}

We used previously published $VRK$ photometry of $\eta$ Aql
\citep{barnes97} to derive its apparent magnitude in the above bands
as a function of phase by fitting a low-order Fourier series to the
published photometry, after first correcting for the effects of
reddening following the procedure outlined in \citet{evans93}. The
individual values of $E(B-V)$ were taken from \citet{fernie90}, and the 
reddening corrections applied are listed in Table \ref{tab:red}.  For
each diameter measurement we then used the Fourier series to derive
$m_V$ and $V-K$ at the epoch of observation, and using Eq. \ref{eqn:sb} we
derived the corresponding surface brightness.  Results are shown in
Figure \ref{fig:fv} and listed in Table \ref{tab:sb}.  We also
performed this type of fit using $\zeta$ Gem data. In this case we
used photometry from \citet{wj68} and \citet{mb84}.

\begin{table}
\begin{center}
\begin{tabular}{cccc}
	   &    &   &   \\
\tableline 
\tableline  
Cepheid	   & $A_V$  & $A_R$  & $A_K$   \\
\tableline 
\tableline 
$\eta$ Aql  & 0.515  & 0.377 & 0.055   \\
$\zeta$ Gem & 0.062  & 0.046 & 0.007   \\
\tableline    
\end{tabular}
\caption{\label{tab:red} Reddening values used in deriving surface
brightness parameters for $\eta$ Aql and $\zeta$ Gem, based on
values of $E(B-V)$ from \citet{fernie90}. }
\end{center}
\end{table}

In Table \ref{tab:sb} we compare the derived surface brightness
relations to similar relations from work based on non-variable
supergiants \citep{fg97} and other Cepheid observations
\citep{nordgren01}.  The $F_{V}$ vs. $V-R$ fits can also be compared with
the \citet{gieren88} result that the slope of the $V-R$ surface
brightness relation (c) is weakly dependent on pulsational period
($P$) according to
\begin{equation}
 c = -0.359 - 0.020 \log{P}
\end{equation}
which for $\eta$ Aql predicts $c = -0.376$ and for $\zeta$ Gem
$c = -0.379$.  
These comparisons reveal generally good agreement between the various
relations in Table \ref{tab:sb}.

\begin{table}
\begin{center}
\begin{tabular}{cccc}
	   &                       &       &        \\
\tableline 
\tableline  
Source                &     a    &   b      & c    \\ 
\tableline 
$\eta$ Aql, this work  &$3.941 \pm 0.005$&$-0.125\pm0.004$&$-0.375\pm0.002$\\ 
$\zeta$ Gem, this work &$3.946 \pm 0.011$&$-0.130\pm0.002$&$-0.378\pm0.003$\\ 
\citet{fg97}           &$3.947 \pm 0.003$&$-0.131\pm0.003$&$-0.380\pm0.003$\\
\citet{nordgren00}     &$3.941 \pm 0.004$&$-0.125\pm0.003$&$-0.368\pm0.007$\\
\tableline    
\end{tabular}
\caption{\label{tab:sb} A comparison between the various surface brightness relations (see text for definitions).
}
\end{center}
\end{table}

\subsection{Period-Radius Relations}

The relation between pulsational period and Cepheid radius has
received considerable attention in the literature, primarily because
early results based on different techniques were discrepant
\citep{fernie84,mb87}. Period-radius relations are also useful in that
they can indicate pulsation mode. This is important for calibrating
period-luminosity relations since different modes will yield different
relations \citep{fc97,nordgren01}.

\begin{figure}[ht]

\plotone{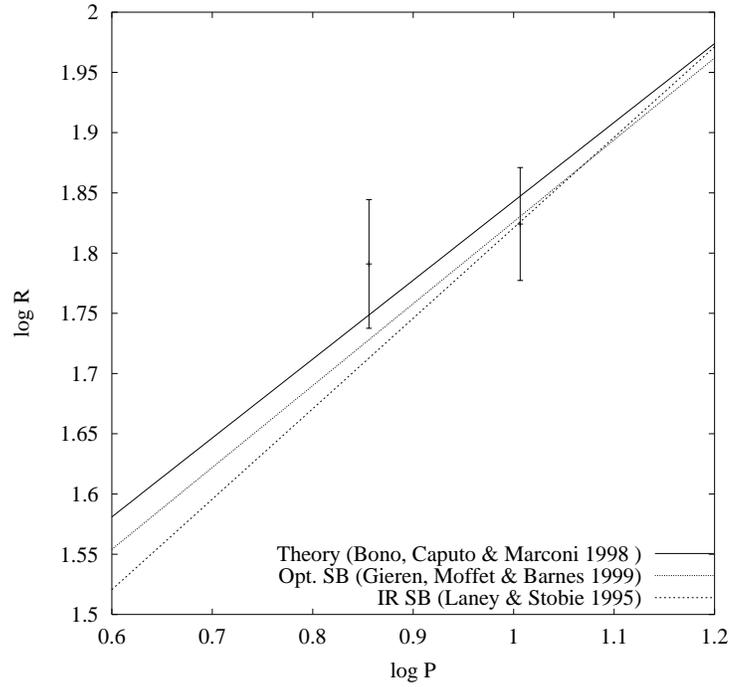}
\caption[Perid-Radius Relations]{\label{fig:rad} Period-radius diagram for the two Cepheids $\eta$ Aql and $\zeta$ Gem, together with three relations available in the literature: a theoretical relation derived by \citet{bono98}, an optical surface brightness relation from \citet{gmb99} and an IR surface brightness relation from \citet{laney95}.  }
\end{figure}

In Fig. \ref{fig:rad} we compare our measured Cepheid diameters to
the values predicted from a range of techniques:
\citet{bono98} calculate a period-radius relation from full-amplitude,
nonlinear, convective models for a range of metallicities and stellar
masses.  \citet{gmb99} use the surface brightness technique based on $V$
and $V-R$ photometry and the \citet{fg97} result to derive radii for 116
Cepheids in the Galaxy and the Magellanic Clouds. They find an
intrinsic width in their relation of $\pm 0.03$ in log R.
\citet{laney95} also use the surface brightness technique for
estimating Cepheid diameters.  However, they find that infrared
photometry ($K, J-K$) is less sensitive to the effects of gravity and
microturbulence (and presumably also reddening), and hence yields more
accurate results.  For shorter periods ($\le 11.8$ days) their results
indicate smaller diameters as compared to other relations.

Given the limited sample of only two radius measurements we can draw
only preliminary conclusions: (1) the general agreement between our
observations and the relations is good, and (2) the data seem to
prefer a shallower slope than e.g.\ the \citet{laney95} relation.
This latter observation will have to be confirmed with observations 
of shorter-period Cepheids.

\section{Summary}

We have measured the changes in angular diameter of two Cepheids,
$\eta$ Aql and $\zeta$ Gem, using PTI. When combined with previously
published radial velocity data we can derive the distance and mean
diameter to the Cepheids. We find $\eta$ Aql to be at a distance of
$320 \pm 32$ pc with a mean radius of $61.8 \pm 7.6 R_{\odot}$.  We
find $\zeta$ Gem to be at a distance of $ 362\pm38$ pc, with a mean
radius of $ 66.7\pm7.2 R_{\odot}$, in good agreement with previous
work. The precision achieved is $\sim$ 10\% in the parameters; further
improvement is at present limited by our understanding of the details
of the Cepheid atmospheres. In particular the details of limb
darkening and projection factors need to be understood, with the
projection factors being the largest source of systematic
uncertainty. 

We note that these results do not rely on photometric surface
brightness relations, hence results derived here can be used to
calibrate such relations. We performed such calibrations and found
good agreement with previous results.  We also note that at present we
have derived distances to only two Cepheids, and although the derived
distances are consistent with currently used period-luminosity
relations, it will be necessary to observe several more Cepheids with
this technique before worthwhile quantitative comparisons can be made.

In the near future long-baseline interferometers will provide a great
deal of useful data in this area: in addition to further observations
of the brightest galactic Cepheids, the very long baselines currently
being commissioned at the Navy Prototype Optical Interferometer
\citep{arm01b} and the Center for High Angular Resolution Astronomy
array \citep{theo01} will allow direct measurements of the limb
darkening effects through observations of fringe visibilities past the
first visibility null. Given the close relation between limb darkening
and projection factors we expect that improvements in understanding
one will improve our understanding of the other.  It is also clear
that additional photometry and radial velocity measurements would be
very useful.  In particular $\zeta$ Gem suffers from a lack of good
infrared photometry, while concerns about level effects make infrared
radial velocity measurements like those of \citet{sasselov90} very
desirable.

\acknowledgements

We thank D. Sasselov, A. F. Boden, M. M. Colavita, S. R. Kulkarni, and
R.R. Thompson for valuable comments.  We also wish to thank K. Rykoski
for his excellent observational work. Observations with PTI are only
made possible through the efforts of the PTI collaboration, for which
we are grateful. Funding for the development of PTI was provided by
NASA under its TOPS (Toward Other Planetary Systems) and ASEPS
(Astronomical Studies of Extrasolar Planetary Systems) programs, and
from the JPL Director's Discretionary Fund.  Ongoing funding has been
provided by NASA through its Origins Program and from the JPL
Directors Research and Development Fund. This work has made use of
software produced by the Interferometry Science Center at the
California Institute of Technology.  This research has made use of the
SIMBAD database, operated at CDS, Strasbourg, France. B.F.L gratefully
acknowledges the support of NASA through the Michelson fellowship
program.

\clearpage

\clearpage

\clearpage

\end{document}